\documentclass{svproc}
\pdfoutput=1

\usepackage{url}
\usepackage{color}
\usepackage{subcaption}
\usepackage{graphicx,dsfont}
\usepackage[utf8]{inputenc}
\usepackage{hyperref}

\begin{document}
\mainmatter              
\title{A Network Science Perspective to Personalized Learning}

\titlerunning{A Network Science Perspective to Personalized Learning}  

\author{Ralucca Gera\inst{1} \and Akrati Saxena\inst{2} \and D'Marie Bartolf\inst{1} \and Simona Tick\inst{1}}
\authorrunning{Ralucca Gera et al.} 
\tocauthor{Ralucca Gera, Akrati Saxena, D'Marie Bartolf, Simona Tick}
\institute{Naval Postgraduate School, Monterey, CA, USA\\
\email{rgera@nps.edu, dmarie.bartolf@nps.edu, sltick@nps.edu}
\and
Eindhoven University of Technology, the Netherlands\\
\email{a.saxena@tue.nl}
}

\maketitle             

\begin{abstract}
The modern educational ecosystem is not one-size fits all. Scholars are accustomed to personalization in their everyday life and expect the same from education systems. Additionally, the COVID-19 pandemic placed us all in an acute teaching and learning laboratory experimentation which now creates expectations of self-paced learning and interactions with  focused educational materials. Consequently, we examine how learning objectives can be achieved through a learning platform that offers content choices and multiple modalities of engagement to support self-paced learning, and propose an approach to personalized education based on network science. This framework brings the attention to learning experiences, rather than teaching experiences, by providing the learner engagement and content choices supported by a network of knowledge, based on and driven by individual skills and goals. We conclude with a discussion of a prototype of such  a learning platform, called CHUNK Learning.

\end{abstract}

\section{Introduction and Motivation}\label{section:introduction} 

Education must meet the changing needs of a complex environment where learners are expected to contribute as creative problem-solvers. Education solutions must satisfy the specific educational needs of each learner while meeting the rapidly changing learning objectives for each degree or job in a resource-efficient way. Consequently, the educational ecosystem must also be a flexible and rich cognitive environment, supported by adaptable high quality content for academic performance as well as social networking for strong emotional support of scholars and complemented by effective learning analytics.

We identify current shortcomings of learning platforms and propose solutions that network science and personalized learning can address.
At the content level, we focus on exploring the relevance of content by anchoring it to each learner's existing knowledge, and how content connects to the skills of each learner.  Additionally, we discuss how interconnecting  people, content, goals, and skills support student learning outcomes and collaborative learning, and what is the impact of these interactions on education experiences.

The goal for a personalized learning model is to facilitate a learning culture that cultivates curiosity and inquiry. While all learners interact with the same  content to show proficiency,  learning differentiation is based on  branching off the main knowledge thread driven by each student's unique attributes (such as experience,  existing skills, and learning goals).  
To achieve such personalization, we propose supporting the learning process by networking the content and the learners, driven by learners' goals and skills they possess.  Then, the network science concepts can be used to identify critical content needed, suggest basic skills, group similar content, recommend learning paths,  identify similar users to support each other in the learning process, and capture and analyze the data  learning patterns to support further improving the learning process.  For example, community detection on the learning content can identify a larger group of interconnected topics to be presented to the learner, both visually as the network of building blocks, as well as descriptively through the content suggested.  Therefore, the learning between topics is not segregated, rather cohesive.


Social network analysis (SNA) of students social networks on learning platforms helps in understanding various phenomenon, such as the group learning behavior of students, the correlation of students' position in social networks and their academic performance, information diffusion among students, the extent of the homophily in the classrooms, and if any student is isolated and needs help~\cite{saxena2019survey}. We can thus extract information straight from the social networking and communication pattern data from the collaborative online learning platform~\cite{Obadi2010FindingPO}. The analysis of such data can improve our understanding of teaching and learning patterns, such as how to increase collaboration among students, what topic to re-teach some topics if not clear to students, identify help-seeking students and provide special assistance and so on. Additionally, this information can improve personalized education by recommending topics to students based on their learning history as well as the learning paths of their friends.

In this work, we propose a personalized education model driven by network science, and present a prototype, called CHUNK Learning, as a solution to the model. We create this agile system by interconnecting the content, skills, and learners, that creatively addresses learning theory while supporting an ecosystem focused on motivating students and improving learning outcomes.  Our focus is on creating a network of knowledge, individualized education pathways, and social network of users to improve personalized education.

\section{State-of-the-art} 

We  present a synthesis of the state-of-the-art research of network science in the support of personalized education, both as a network of knowledge as well as the social network of users, focused on the educational ecosystem.

\subsection{The Design of a Network of Knowledge}

Traditional education is linear, one chapter after another through a whole course, or course after course through a degree. Learners engage in some courses with the same classmates and some courses with new classmates, creating a social network that is not usually captured and capitalized. An interconnected model of education brings  an interconnected (non-linear) view of both the knowledge and the social network. That is, we can model the learning as a user navigating through a network of knowledge built based on an ontology, or pre-requisites, or dependency in some other way. Additionally, we model the learning support as the social network of the users, namely, content creators, instructors, instructional support staff, stakeholders, sponsors, and learners.  Then apply network science to improve both teaching and learning.

Recent research identifies the benefit of interconnecting the knowledge for the learners, seeking models for a networks of knowledge.  For example, one way to model it is by ``quantifying and analyzing the structure of students’ knowledge of a given discipline as a knowledge network of interconnected concepts". Once such model is created, then we can capture the learning and the retrieval of the information that learners interacted with over time~\cite{cleven2018multilayer,siew2020applications}. 

\begin{figure}
    \centering
    \includegraphics[width=1.0\linewidth]{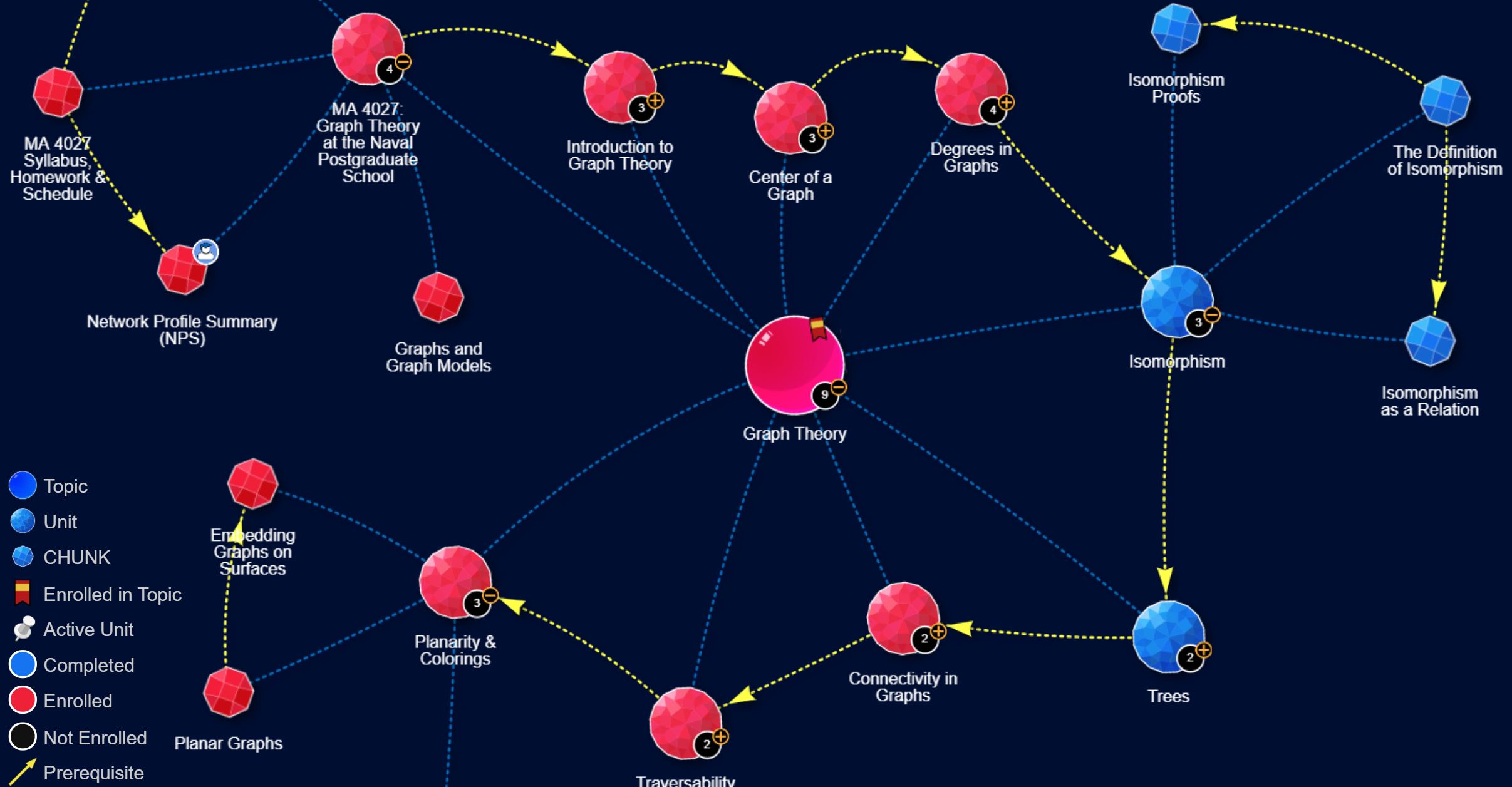}
    \caption{A portion of a network of digestible micro-lectures \protect\cite{CHUNKLearningnet}.} 
    \label{fig:FundamentalsOfMath}
\end{figure}

In the interconnected world of 21st century education,  subject matter experts can collaborate to create a network of knowledge as a curated collection of networked learning modules, organized to meet strategic objectives (academic or professional). Gera et al. \cite{CHUNKLearningnet,gera2019chunk} successfully created a prototype of this in 2018 through the Curated Heuristic Using a Network of Knowledge for Continuum of Learning (CHUNK  Learning) concept, a real-time and adaptive teaching-learning modular method for enhanced and personalized education. Figure~\ref{fig:FundamentalsOfMath} shows a view of the portion of such network of chunked content into connected digestible micro-lectures captured by the red nodes.  Explorations of this network to support education shortly followed, by looking at a network science approach  to recommending learning paths for students~\cite{andriulli2019adaptive,diaz2019recommender,gutzler2019adaptive}. These innovative ways of educating bring the attention to the network of ideas, and how understanding this network supports student learning success.

\subsection{The Education Pathways Supported by the Networks}

Cognitive flexibility theory discusses learning within complex and ill-structured knowledge domains, through the inability of linear educational structures to support meaningful learning experiences~\cite{boger1996cognitive}. In these environments the application of network science demonstrates complex concepts' interconnections and supports semantic memory.  Additionally, the use of nodes and edges creates visual learning pathways that provide an avenue for supporting and tracking learners' knowledge acquisition~\cite{siew2020applications}.   

The use of network science to create a structural representation of learning paths supporting knowledge acquisition allows educators to determine if there is an optimal path.  Networked education pathways provide the necessary environment to examine average length of paths between two nodes in the network and the relationship to student understanding of material~\cite{siew2019using}. This also presents an opportunity to examine the use of bridging concepts within a complex knowledge structure~\cite{siew2020applications}.  In this way, a feedback mechanisms exist to inform educators about how to improve education experiences for learners, filling in gaps within the knowledge structure as needed in order to support desired learning outcomes.

\subsection{Incorporating Online Social Networking in Learning Platforms}
 


Online learning platforms complement or replace classroom learning for students; however, they usually lack students engagement. One way to support the students engagement in such educational platforms is by incorporating social networking into the learning platform. 
Researchers have studied the benefits of networking the learners in learning and professional development~\cite{saxena2019survey}.  

Recently, various existing platforms 
provide the ability for students to communicate with each other from all over the world. While this is a wonderful step in supporting students, these platforms lack social networking among student, such as building a friendship network, or recommending for collaboration. To our knowledge, no existing platform uses the social networking data for personalizing education, such as recommending  content or suggesting teams to test or work on group projects. CHUNK Learning includes the vision for such a platform~\cite{CHUNKLearningnet}.

We seek to introduce educational platforms that are empowered by the social networking to support meaningful learning timely and respectful of learner’s time in a cost-effective manner. 
Research on the CHUNK Learning platform considered the social networks to recommend similar content to similar learners, as well as connecting similar learners based on their learning paths, background, and mentor-mentee type relationships~\cite{gutzler2019adaptive}. 
\section{Evidence-based Approach for Personalized Education} \label{section:evidence} 
Learner-centered  educational  environments have the capability to adjust the learning to meet the needs of individuals and groups of learners, recognizing that the needs and motivators may be different. According to the U.S. Department of Education, personalized learning looks to place the learner at the center of the environment in hopes of addressing current challenges in education~\cite{thomas2016future}. The introduction of technology has enhanced the ability to support this type of environment at scale. Education data mining is meant to explore large-scale data from learning environment interactions to develop recommendation systems, predict student's performance, group students according to their similarities, detect undesirable behaviour~\cite{george2019review}.

Personalization of learning helps to address the challenges resultant of insufficient student-support during the learning process, content perceived as irrelevant, and decreased learner motivation as a result of feeling disconnected with the learning process and the course material~\cite{george2019review}. It addresses learners’ current competency level, prior knowledge on the subject or related subjects, and learning style preferences~\cite{bernard2017learning} through the use of a learner profile. Studies show that personalized learning environments support improved learning effectiveness by meeting the needs of each learner through improved learning outcome achievement, learner satisfaction with the learning process, and determination of self-efficacy~\cite{xu2014enhancing}.  These studies also demonstrate the connection between learner perception and actual achievement.  When the learner had a higher degree of self-efficacy, the learning experience is improved, and thus the learning performance is improved. 

Personalized learning takes into consideration cognitive, metacognitive, motivational, and social/emotional competencies~\cite{redding2014personal}.  It also addresses the learners’ desire for meaningfulness and relevance, and learning that is interest-driven and self-initiated. In this way, it focuses on context personalization learning theory~\cite{doi:10.1080/15391523.2020.1747757} and the ability to align content to the interests of the learners.  Studies have shown that content relevance is directly linked to learner retention~\cite{hone2016exploring}. By increasing the interest of learners, personalized learning helps ensure learners will remain engaged and supports deep-processing learning~\cite{walkington2014motivating}.  

When students are given choices about their learning, in a learning environment that is flexible and responsive to student needs, learning outcomes are shown to improve.  Additionally, they do so at a faster pace for learners who rank lower in achievement in a traditional learning environment~\cite{RR-1365-BMGF}. There is a large variety of personalized learning environment features that have been shown in the literature to be effective, from those building flexibility and choice in interaction with the context via blended learning, to enabling students to personalize in a meaningful way the learning content~\cite{doi:10.1080/15391523.2020.1734507}. 

\section{A Vision for Personalization in Educational Ecosystem} \label{section:proposal}

We now  propose  a network framework for content, based on relationships between contents' tags, and  learners' skills and preferences,.  We provide choices for learners on how and when to engage with content, presenting personalized education plans that build on each  leaner’s existing skills.
We now present how such a network science based framework can support personalized education, and conclude with a prototype used by the authors for the last several years.

\subsection{Pathways for Personalized Learning}

Education is improving, but not as fast as our aspirations.  Learners  bring distinct backgrounds,  learning styles preferences, and different motivations for engaging with the content.  This surfaces the need for personalized learning that assesses each the learner's gaps, skills and prior experience before bringing in new information, thus preventing more gaps creation or repeat of lessons already learned.  

It also requires that a variety of learning styles be available for learners when engaging with  instructor or the asynchronous content. Content and applications of the newly learned content must differ  based on each learner's experience and background  (learning and using a mathematical concept for an economist versus a mathematician is different).  Content must have applications that are relevant to the learner's background to anchor it to learner's experience, since they will be applying it in the future. 

Additionally, some learners are directed to learn specific content for certification as needed by a job or degree, while the life long learners explore the network of knowledge for personal indulgence or personal desire to up-skill their existing expertise. Lastly, the personalization needs to engage every learner based on his/her own interests to anchoring to the learner's personal experiences to promote active learning. 
With the goal of providing a personalized experience to each student, choices of learning paths for each students need to be dynamically created. 
A method to create these learning paths can utilize network science and data mining techniques on the annotated network of content and users~\cite{cleven2018multilayer}.  

Once users have navigated even a portion of the network of knowledge, their experiences create a database of learning paths, which can be analyzed to understand the learning patterns based on gender, age, backgrounds, learner type (directed versus exploratory learner), or other attributes of the learners~\cite{lucero2020optimizing}. These findings can be used to further improve the personalized learning platform by suggesting what content should be added and designing a better recommendation system for learners. 



\subsection{Engaging with other Users}

A social network complements and supports the network of content as a support element of the learning process.
The learning platform should provide the service to maintain the social connections through a social network integrating learners and content in the learning platforms in two ways. First, users can follow or be friend with people they know or they want to connect with for personal interest. Secondly, the platform should suggest new users to connect with based on similar background and same learning interest. Consequently, the social network data can be mined by the recommender of the Network of Knowledge to suggest new relevant content to users. 

Additionally, to incorporate group learning, the platform should suggest the teams either for instant collaborating, or on group projects based on the current learning path.  Also, the recommendation system must adaptive in recommending such teaming patterns based on past performance of the students using various learning techniques and interests. The grading or rewards  must be defined based on the overall learning of the students, as well as the individual learning that could be assess in advance. Subsequently, the social network supports the learner's journey both in the learning process as well as in the assessment.


\subsection{A Working Example: CHUNK Learning Platform}

A prototype of personalized education based on network science is CHUNK (Curated Heuristic Using a Network of Knowledge) Learning~\cite{CHUNKLearningnet} platform described in details by Gera et al. in~\cite{gera2019chunk}.  We now point out how CHUNK Learning fulfills some of the identified needs for a network science based personalized education. Since the existing CHUNK Learning system  is in its infancy state right now, we will conclude this subsection with ideas that can extend the current version based on the discussions of Sections~\ref{section:evidence} and~\ref{section:proposal}.
  
The CHUNK Learning prototype offers several features identified in this work: (1) personalized learning journey by using information provided in the learner’s profile to automatically recommend the most relevant-to-you content, (2) a content mapping illustrates how a learner can progress through the network of knowledge, to include the choices of learning paths through the network, and (3) framework to support a social network of learners based on their profile and content they are engaging with at each time they log into CHUNK Learning.

CHUNK Learning breaks away from the linear traditional education models and provides content delivery that respects and builds on learners' different skills, knowledge, learning styles, and approaches to problem-solving. Learners are empowered by a system that ensures the learning journey is cohesive, flexible, and respectful of learner's time and interests, seamlessly  infusing  directions from an administrator/supervisor. This personalized education is accomplished through the use of the following components.
\begin{itemize}
    \item Content is chunked into intense, short, and focused educational modules (chunks) that are interconnected in the network of knowledge to provide (a) learner's choices and platform's recommendation to build personalized learning paths, (b) view of the learning path choices through the content, and (c) the big picture of cohesive learning.
    \item Choices of interchangeable and reusable content stimulate each learner's interest and provides relevance of each topic to each learner~\cite{gera2019chunk}.
    \item Personalized content driven by  implicit recommendations from a user's social network based on tie strength between 
    learners~\cite{critchley2021recommender}.
    \item Learners' are placed into a social network to support each-other, such as pairing senior learners as mentors for junior ones, based on their automatically updated profiles~\cite{reeder2021analysis}. 
\end{itemize}

Based on learner's profile, CHUNK Learning optimizes both the content and the methodology delivery  to meet the needs of each learner. It is intended to fit typical academic needs, much like curriculum mapping, while personalizing the content within the chunk of knowledge. That is, content is organized hierarchically using Topics, Units, Chunks, and Chunklets to index content in a way to meet and improve the overall coherence of a course of study (more information on CHUNK Learning Wiki~\cite{wiki}. 

\begin{figure*}[t]
    \centering
    \includegraphics[width = .95\textwidth, keepaspectratio]{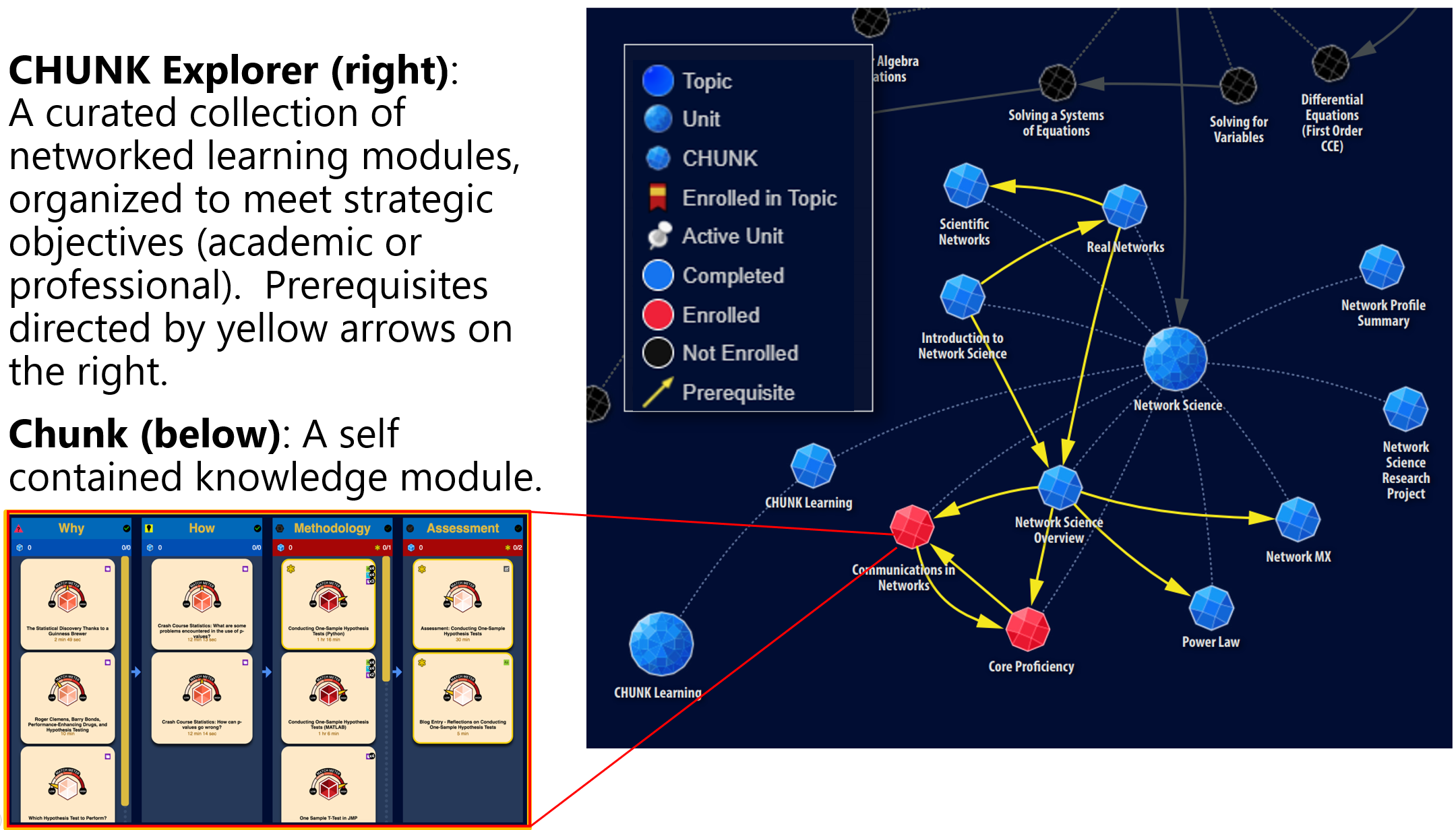}
    \caption{The network explorer and a chunk's content in CHUNK Learning~\cite{CHUNKLearningnet} }
    \label{fig:CHUNKExplorerAndChunks}
\end{figure*}

Figure~\ref{fig:CHUNKExplorerAndChunks} shows the personalized adaptive learning framework by displaying a portion of the network of knowledge on the right, the personalized content of a chunk of knowledge on the bottom left of the figure.  Within a chunk, the top row of content reveals system's recommendations based on learner's profile, and the other tiles are the choices of the ranked alternatives.
The system displays content in a manner that can be viewed in a network format through the CHUNK Learning explorer at the topic, unit, and chunk level.


The explorer view at the topic-unit-chunk level, is a concept mapping that follows a logical order to provide each learner a well-rounded and comprehensive educational experience, with choices for deeper dives and connections to other topics-unit-chunk. At the chunklet level, a learner can see the choices for all the building blocks needed to attain any specific academic competency while being recommended the most relevant one based on that learner's profile. Each column of a chunk provides the recommended chunklet at the top, followed by rank ordered choices to complete that part of the chunk before heading to the assessment.  

\begin{figure*}[t]
    \centering
    \includegraphics[width = .95\textwidth, keepaspectratio]{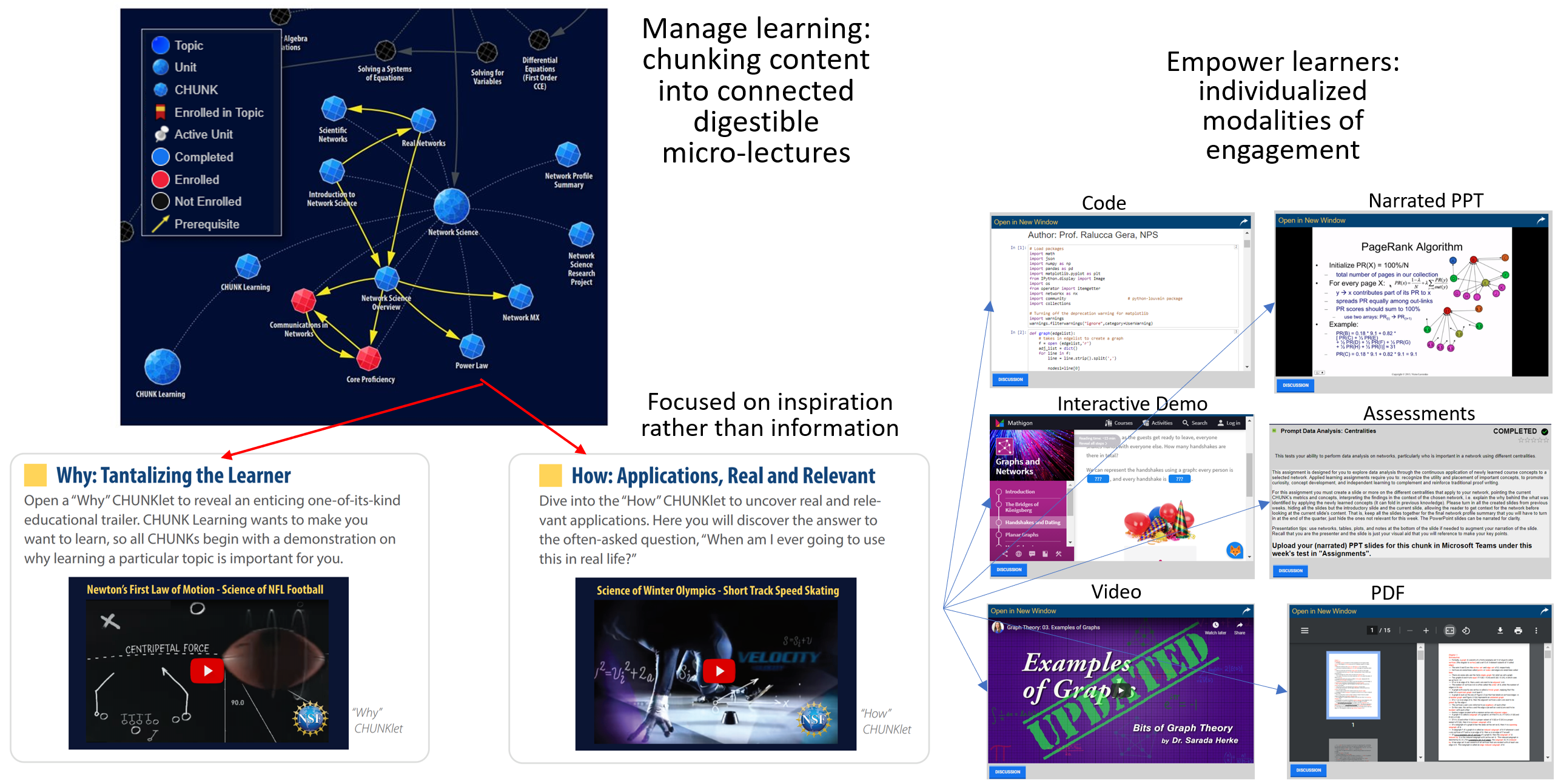}
    \caption{A view of personalized learning pathway  in CHUNK Learning~\cite{CHUNKLearningnet}}
    \label{fig:PersonalizedLearningPathway}
\end{figure*}

Figure~\ref{fig:PersonalizedLearningPathway} shows an example of  personalization in CHUNK Learning achieved through (a)  interchangeable chunklets  that personalize the application of the content  (on the left side of the figure) and (b)   personalized  interchangeable activities  that speak to the different styles of learning, such as video, PDF, code, demo, interactive activity (on the right side of the figure).

CHUNK Learning is a prototype used  for hybrid teaching at the Naval Postgraduate School, CA, USA, by Prof. Ralucca Gera (2018- current), and by Dr. Michelle Isenhour (2018-2019).  For these classes, students learn part of the content asynchronously using CHUNK Learning,  complemented by synchronous practice and discussions. The assessment of students' experience in these classes shows promising benefits in using CHUNK Learning for hybrid teaching. 
Besides the self-paced environment, students valued the mix of modalities of engagement with the content available in CHUNK Learning  supporting  meaningful engaging interactions in class, while enhancing their learning experience.  

\subsubsection{Extending the CHUNK Learning Platform.}

How can this prototype be extended to build further on the field of network science? We see four main areas where we can applying network science, and some progress has already been published in each of the areas: (1) the network of knowledge's explorer view, (2) the content withing a chunk, (3) the social network of users, and (4) to create a network driven recommendation system and learning analytics. Below we expand on the existing work on each of these extensions. 

Currently, the explorer view of the Network of Knowledge in CHUNK Learning is static and identical for all users (except the color coding based on the enrolled and completed chunks).  
To focus and place  the learner on a meaningful learning path, we propose a personalized display of the network of knowledge  based on assessment and relevancy to user's learning objectives.  

Complementing the personalizing of the explorer view, we can personalize each chunk's content. Currently it is performed using the content's tags compared to the learner profile's tags, in addition to the ranking of the content based on the user's rating of each chunklet.  The two existing rankings are based on quality of content (1-5 starts) and based on the relevance to the learner (thumbs up or down).

For the social network approach, we focus on a dynamic social network of students based on social and academic attributes. Learners can compete or support each other by pairing up for interactive activities on the spot, junior members can be matched to their seniors to learn from their experience to complement the theory~\cite{reeder2021analysis}. Additionally, 
the recommender system  can then exploit the connections between users and content~\cite{critchley2021recommender}.  

Currently, the recommender system  updates the learning plan based on each learner's activities (learned, viewed, tested), keyword searches, and content ratings by increasing or reducing the strength of the connection between the learner profile and activities. Extensions have been proposed using link prediction, community detection, and partial information from the network's tags~\cite{andriulli2019adaptive,diaz2019recommender,gutzler2019adaptive}.

\subsection{Challenges of Network Science based Personalized Education} 
Personalized learning and incorporation of network science for a personalized learning environment has its challenges. In many ways, the proposed environment requires changes in design of curricula and a shift in an education culture.
To support that culture, training is required to ensure content creation, structure, and tagging create a coherent learning experience~\cite{zheng2018digital}. Additionally, for  personalizing content by providing content in different modalities, we need to ensure that content authors are using proper instructional design methods. 

A technology-rich education environment increases the variety of digital resources used to support multimodal learning with a challenge of data collection and meaningful data analytics~\cite{bingham2018ahead}.  Interpretability of data is necessary to support proper implementation and scalability. Data-driven decision making is at the core of successful personalization 
ensuring the platform remains agile and adaptive.  Additionally, it is imperative that ethical concerns are identified and addressed to include information privacy, surveillance, safe-guarding from discrimination, and for research and experimentation~\cite{regan2019ethical}.



Personalized learning receives increased attention from education scholars. While the theoretical contributions from several disciplines within learning sciences abound, the empirical studies that put to test ideas of  personalized learning are few and often present mixed results~\cite{doi:10.1080/15391523.2020.1747757}. 
Large-scale randomized controlled studies are needed to provide evidence for how well and in what context personalized education can improve learning outcomes and experiences as well as the efficacy of implementation~\cite{RR-1365-BMGF}. While the efforts to use learning analytics have recently intensified, the evidence presented by recent studies are from smaller-scale quantitative, mixed-methods, or qualitative studies which lack the power afforded to large-scale randomized controlled studies to validate findings on the efficacy of personalized learning~\cite{doi:10.1080/15391523.2020.1734507}. To be able to test the efficacy of personalized learning we need to improve the data collection efforts and the analysis methods to examine the merits of personalized learning. Network science can have an important role in the learning analytics effort, with its ability to show relational data and the learning path for each learner. 

\section{Conclusion}\label{sec:conclusions} 

In this paper, we briefly discussed how the network science can help improve  personalized education at scale, and we introduced a prototype named CHUNK Learning. Based on our vision, the personalization of the prototype is driven by concepts of network science applied to  the network of knowledge and the content recommender system. Further, it can be extended to include social network of users in support of personalized learning. We discussed additional  extensions of the use of network science, and conclude with some challenges to consider in creating and using such a personalized learning ecosystem.  We share this work to promote conversations and research towards a possibly new field of personalized education driven by network science that we will further explore.

\bibliographystyle{unsrt}
\bibliography{citations}

\end{document}